\documentclass[aps,prb,twocolumn,nofootinbib]{revtex4}  
\usepackage{amsmath,amsfonts,amssymb,latexsym}
\usepackage{hhline}
\usepackage{graphicx}
\usepackage[arrow,matrix]{xy}
\usepackage{tikz}
\usepackage{tikz-cd}
\usetikzlibrary{positioning,arrows}
\usetikzlibrary{decorations.pathmorphing}
\usetikzlibrary{decorations.markings}
\newcommand{\tp}{\otimes}
\newcommand{\ra}{\rightarrow}
\newcommand{\cob}{d} 

\newcommand{\zz}{\mathbb{Z}}

\newcommand{\mcg}{\mathcal{G}}

\newcommand{\mcp}{\mathcal{P}}

\newcommand{\zt}{\mathbb{Z}_2}

\newcommand\be            {\begin{equation}}
\newcommand\ee            {\end{equation}}
\newcommand\ba            {\begin{aligned}}
\newcommand\ea            {\end{aligned}}

\newcommand{\mcl}{\mathcal{L}}

\newcommand{\mco}{\mathcal{O}}

\newcommand{\mfw}{\mathfrak{w}}

\newcommand{\rep}{\text{Rep}}

\newcommand{\ztp}{\mathbb{Z}_2^\mathcal{P}}
\newcommand{\ztt}{\mathbb{Z}_2^\mathcal{T}}
\newcommand{\p}{\partial}

\newcommand{\cupp}{\wedge}
\usepackage{verbatim}
\begin{document}

\title{Anomalies and symmetry fractionalization in reflection-symmetric topological order}  
\author{Ethan Lake}
\email{lake@physics.utah.edu}
\affiliation{Department of Physics and Astronomy, University of Utah, Salt Lake City, UT 84112, USA}

\date{\today}

\begin{abstract} 
One of the central ideas regarding anomalies in topological phases of matter is that they imply the existence of higher-dimensional physics, with an anomaly in a D-dimensional theory typically being cancelled by a bulk (D+1)-dimensional symmetry-protected topological phase (SPT). 
We demonstrate that for some topological phases with reflection symmetry, anomalies may actually be cancelled by a D-dimensional SPT, provided that it comes embedded in an otherwise trivial (D+1)-dimensional bulk. We illustrate this for the example of $\zz_N$ topological order enriched with reflection symmetry in (2+1)D, and along the way establish a classification of anomalous reflection symmetry fractionalization patterns. In particular, we show that anomalies occur if and only if {\it both} electric and magnetic quasiparticle excitations possess nontrivial fractional reflection quantum numbers. 
\end{abstract}

\maketitle  

\section{Introduction}

A major research direction in recent years has been the study of the interplay between symmetry and topology in quantum phases of matter. When a gapped topological phase of matter possesses a symmetry, its quasiparticle excitations can carry fractional quantum numbers of the symmetry group, with common examples including the fractional electric charge carried by the quasiparticle excitations in the fractional quantum Hall states\cite{laughlin1983anomalous} and the $S= 1/2$ spinon excitations in quantum spin liquids \cite{anderson1987the,savary2016quantum}. The classification of the different ways in which symmetry fractionalization can occur is both an interesting theoretical problem and a valuable question from an experimental standpoint, as symmetry fractionalization can be measured by conventional shot-noise and neutron-scattering experiments \cite{de1998direct, tennant1993unbound, essin2014spectroscopic}, in contrast to other signatures of topological order like anyonic braiding statistics, which are notoriously difficult to measure. 

Closely related to the idea of symmetry fractionalization is the study of anomalies in topological phases. Anomalies in topological phases arise when there is an obstruction to gauging a symmetry, and their presence depends on the pattern of symmetry fractionalization realized in the system. Anomalies in D spacetime dimensions must typically be cancelled by (D+1)-dimensional physics\cite{kapustin2014anomalousprl, cho2014conflicting, dijkgraaf1990topological}, which usually comes in the form of a symmetry protected topological phase\cite{chen2015anomalous, kapustin2014anomalies, wen2013classifying, barkeshli2014symmetry, burnell2014exactly, vishwanath2014physics,wang2016bulk, fidkowski2013nonabelian, hung2013quantized} (SPT), which is a short-ranged entangled phase possessing a unique ground state on arbitrary spatial manifolds. \cite{gu2009tensor,schuch2011classifying,chen2013symmetry,chen2011classification,chen2016symmetry}. 

The majority of work devoted to understanding fractionalization and anomalies in gapped symmetry-enriched topological phases (SETs) has focused on the case
where the topological phase in question possesses an internal (onsite) unitary symmetry. Although there has been some recent work examining time reversal and space group symmetries\cite{qi2015anomalous, cheng2016exactly, essin2013classifying,kapustin2014symmetry,thorngren2015framed,hermele2015bosonic}, a systematic understanding of how symmetry fractionalization and anomalies occur in these systems is still under development. Developing such an understanding is particularly important from an experimental point of view, due to the abundance of these types of symmetries in real materials.


In this paper, we use a dimensional reduction procedure to show that for abelian SETs enriched by $\ztp$ reflection symmetry, anomalies in D spacetime dimensions may be cancelled by a D-dimensional SPT, as long as it is embedded in an otherwise trivial ambient (D+1)-dimensional bulk. We focus on the illustrative example of $\zz_N$ topological order, and use a field theory approach to classify anomalous $\ztp$ fractionalization patterns, which agrees with recent results obtained for the case of $\zz_2$ topological order\cite{qi2015anomalous,song2016pgspts}. We show that in these theories, anomalies arise if and only if both the electric and magnetic quasiparticles carry a nontrivial fractional reflection quantum number and that when they do occur, anomalies can be understood by considering the way in which gauge fields for the $\ztp$ symmetry, which we define on reflection-invariant subspaces of the system, interact with one another. 


\section{Model setup and dimensional reduction procedure} \label{sec:model}

\begin{figure}
\includegraphics[scale=1]{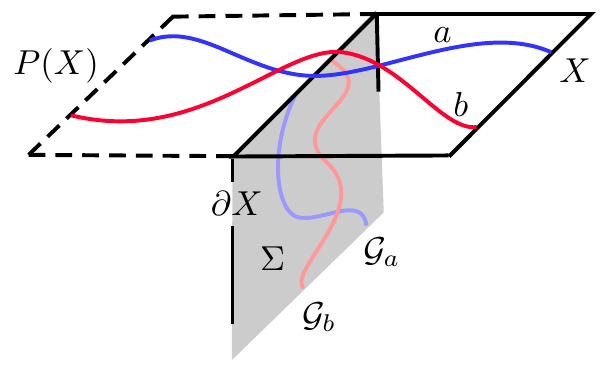}
\caption{\label{fig:strings_geometry} An illustration of the geometry we consider and the gauge fields present in our theory, viewed at a single time slice. The $a$ and $b$ cochains represent electric and magnetic quasiparticle strings, and live on the surface $X$, although our theory only involves their restriction to the mirror axis $\p X $. The $\mcg_\alpha$ represent the electric and magnetic $\ztp$ gauge fields, and are initially also defined on $\p X$. Deciding whether or not the $\mcg_\alpha$ also must extend down into the mirror plane $\Sigma$ (as shown in the figure) amounts to determining whether or not the surface theory is anomalous. }
\end{figure}

In this paper, we will restrict our attention to topological order enriched with $\ztp$ reflection symmetry. In crystalline solids translation symmetry will always exist alongside reflection symmetry, which we discuss briefly in section \ref{sec:translation}. We consider a topological phase defined on a (2+1)D surface split along a reflection axis into a manifold $X$ and its image under reflection $\mcp(X)$, with the reflection axis corresponding to $\p X$. Depending on whether or not the theory is anomalous, it may be necessary to place this surface on top of a (3+1)D bulk, with the full system possessing reflection symmetry about a mirror plane $\Sigma$ (see Figure \ref{fig:strings_geometry}).

We assume that the (2+1)D surface $X$ is equipped with a triangulation and for concreteness that it possesses $\zz_N$ topological order, although the generalization to arbitrary abelian topological phases is straightforward. Inspired by Refs \cite{kapustin2014symmetry,thorngren2015framed,wang2015field}, we describe the $\zz_N$ topological order by electric and magnetic {\it integer-valued} 1-cochains $a$ and $b$, which can be thought of as the discrete analogues of the two $U(1)$ fields employed in the conventional $K$-matrix description of $\zz_N$ topological order\cite{wen2004quantum}. We take the electric 1-cochain $a$ to be defined on the links of $X$ and the magnetic cochain $b$ to be defined on the links of $X^*$ (the dual triangulation of $X$), so that the electric and magnetic strings always intersect each other transversely. By Poincare duality, $a$ ($b$) is dual to a two-dimensional submanifold of $X$ ($X^*$), which represents the worldsheets swept out by the electric (magnetic) quasiparticle strings. The coboundary $\cob a$ ($\cob b$) is dual to the boundaries of these worldsheets, which represent the worldlines of the electric (magnetic) excitations in the theory. 

The electric and magnetic excitations interact topologically with each other by passing through each other's string worldsheets, and so the surface theory needs to keep track of how electric quasiparticle worldlines (described by $\cob a$) intersect magnetic string worldsheets (described by $b$) and vice versa. We describe such an interaction on $X$ by a Chern-Simons-like surface action\cite{kapustin2014symmetry,thorngren2015higher}
\be \label{eq:SX} S_{X} = \frac{1}{2N} \int_X (a\cupp \cob b + b\cupp \cob a),\ee
where the prefactor ensures that only values of the integer-valued gauge fields $a$ and $b$ mod $N$ are physical.

To ensure reflection symmetry we take $S_{\mcp(X)} = \mcp(S_X)$ on the other half of the surface (see figure \ref{fig:strings_geometry}). In the expression for \eqref{eq:SX}, we have abused notation slightly and written $\cupp$ for the cup product, which is the discrete analogue of the wedge product. 
In geometric terms, the term $a\cupp\cob b$ represents the intersection of the submanifolds defined by the Poincare duals of $a$ and $\cob b$. More precisely, if we let $a^*$ and $(\cob b)^*$ denote the Poincare duals of $a$ and $\cob b$, which are respectively two-dimensional and one-dimensional submanifolds, then the cup product $a\cupp \cob b$ represents the intersection $a^* \cap (\cob b)^*$. For a review of the essential algebraic topology we refer the reader to Appendix \ref{sec:algtop}. 

One way to motivate our approach to classifying symmetry fractionalization patterns and identifying anomalies is inspired by the dimensional reduction classification of SPT phases 
protected by point group symmetries recently introduced in Ref\cite{song2016pgspts}. We rule out 
a topologically ordered (3+1)D bulk by assumption and assume that $\ztp$ is the only symmetry present. This means that if an anomaly in the surface theory forces any nontrivial physics to occur in the bulk, the bulk theory will necessarily possess only short-ranged entanglement, which may or may not be protected by reflection symmetry. 

Following Ref\cite{song2016pgspts}, we
consider a small volume $V$ lying in the bulk. Because of the short-ranged nature of the bulk entanglement, we may act on the bulk with local unitary operators $U_V$ and $U_{\mcp(V)}$ supported on the volume $V$ and its image under reflection $\mcp(V)$, which act to transform the Hamiltonian densities $\mathcal{H}_V$ and $\mathcal{H}_{\mcp(V)}$ into those of a product state in a reflection-symmetric way. This works for any volume $V$ in the bulk provided that $V$ is not mapped to itself under reflection. We see that by repeatedly applying such $U_V U_{\mcp(V)}$ operators we can disentangle the entire bulk into a trivial product state, with the exception of the mirror plane $\Sigma$ which is mapped to itself under reflection. Thus, we expect that any anomaly-induced bulk physics will derive solely from what happens on the mirror plane, with the rest of the bulk playing a passive role. Now, deciding whether or not the surface topological phase is anomalous is equivalent to determining whether or not the theory can be defined 
solely on the surface $X$, with the {\it entire} bulk (including the mirror plane) remaining in a trivial product state. Since our disentangling procedure means that only the mirror plane can possibly contain any nontrivial bulk physics, we expect that the physics of the interplay between the $\ztp$ symmetry and the surface topological order will be encoded entirely in the boundary of the mirror plane, which is the (1+1)D mirror axis $\p X$, on which $\ztp$ acts in an {\it onsite} way. This means that we have dimensionally reduced the study of a (2+1)D theory to a simpler (1+1)D problem. This dimensional reduction approach can also be motivated by folding the (2+1)D surface about the mirror axis, the details of which we sketch in Appendix \ref{sec:folding}. 

This approach suggests that the symmetry fractionalization on the surface is realized ``holographically'', in the sense that the entire information of the symmetry action is encoded in the properties of the one-dimensional mirror axis. 
Our goal is to construct a field theory description of the mirror axis $\p X$ with gauged $\ztp$ symmetry, which will lead us to a classification of the possible symmetry fractionalization patterns and anomalies that can occur.

\section{Testing for anomalies and classifying fractionalization patterns} \label{sec:anomalies}

The usual way to test for anomalies in SETs is to attempt to gauge the symmetry \cite{etingof2009fusion,kapustin2014anomalousprl,cho2014conflicting}. Normally gauging a non-local symmetry like reflection would be physically ill-defined, although our dimensional reduction approach tells us that we only need to gauge $\ztp$ on the mirror axis $\p X$, on which it acts as an onsite $\zz_2$ symmetry. When we gauge $\ztp$ on $\p X$, we introduce symmetry fluxes for the $\ztp$ symmetry which proliferate on the mirror axis. Symmetry fluxes act on the $a$ and $b$ strings which pass over them, and so in order to fully gauge $\ztp$ on $\p X$, we need to give the $a$ and $b$ $\zz_N$ gauge fields a way of commuting with the action of the symmetry fluxes. This is done by constructing an action which ``intertwines'' the action of $\ztp$ on the $a$ and $b$ fields in a local way. 

More precisely, the gauging procedure is accomplished by introducing two $\zz_2$ gauge fields $\mcg_a$ and $\mcg_b$ which are 1-cochains defined on $\p X$ and $(\p X)^*$, respectively. $\mcg_a$ and $\mcg_b$ are each closed modulo 2, and are the connections of the $\ztp$ gauge field on the mirror axis for the electric and magnetic sectors of the $\zz_N$ topological order. 
To write down a potential field theory description of the mirror axis after $\ztp$ has been gauged, we need to couple the $\zz_N$ gauge fields to the $\mcg_\alpha$ gauge fields. This is done by coupling the electric cochain $a$ to the {\it magnetic} gauge field $\mcg_b$, and the magnetic cochain $b$ to the {\it electric} gauge field $\mcg_a$. Since $\mcg_a$ $(\mcg_b)$ is defined on $\p X$ ($(\p X)^*)$, this coupling ensures that the $\zz_N$ gauge fields always intersect the $\mcg_\alpha$ to which they couple transversely. 

The strengths of the couplings between the $\zz_N$ and $\mcg_\alpha$ gauge fields are determined by the $\ztp$ gauge charges of the $a$ and $b$ fields. These gauge charges can be identified with the reflection quantum numbers carried by the electric and magnetic quasiparticles, which determine the symmetry fractionalization class of the theory. Since reflection acts in an onsite way on the mirror axis, the different symmetry fractionalization classes are parametrized by choices of group cohomology classes $\mathfrak{w}_{\alpha} \in H^2_\mcp(\zt,\zz_N)$ for $\alpha \in \{ a,b\}$ \cite{kitaev2006anyons,barkeshli2014symmetry}, where the cohomology group is twisted by the action of reflection on the $\zz_N$ gauge fields. We will adopt the notation $\omega_\alpha = \mfw_\alpha(\mcp,\mcp)$, so that the single quasiparticle state $|\alpha\rangle$ transforms under reflection as 
\be \mcp:|\alpha\rangle \mapsto e^{\pi i \omega_\alpha/N} |\mcp(\alpha)\rangle,\ee meaning that $\omega_{\alpha}$ determines the reflection quantum number ({\it i.e.} the $\ztp$ gauge charge) of $|\alpha\rangle$. 

Quantifying the topological interaction between the $\zz_N$ and the $\mcg_\alpha$ fields is accomplished by taking the cup product of the electric $\zz_N$ gauge field $a$ with the magnetic $\ztp$ gauge field $\mcg_b$, and vice versa. This means that gauging $\ztp$ on the mirror axis results in the mirror axis action
\be\label{eq:spx}\ba & S_{\p X} = \frac{1}{2N} \int_{\p X} ( \omega_b\mcg_b\cupp a  +  \omega_a\mcg_a\cupp b), \ea\ee
where $a$ and $b$ in the above action are taken to mean the restriction of the surface $a$ and $b$ strings to the mirror axis $\p X$, which by Poincare duality may be viewed as the worldlines drawn out by the intersections of the surface $a$ and $b$ strings with the mirror axis. 

Importantly, the $1/2N$ prefactor in the mirror axis action $S_{\p X}$ means that if the reflection quantum numbers determined by the $\omega_\alpha$ are nontrivial, the mirror axis can effectively possess a sort of $\zz_{2N}$ gauge theory, rather than a $\zz_N$ gauge theory. This is actually quite natural when viewed from an algebraic point of view. In algebraic language, the ways in which reflection symmetry can fractionalize for the quasiparticles in the electric and magnetic sectors are classified by the different group extensions of $\ztp$ by $\zz_N$, which are given by groups $E$ with $E/ \zz_N \cong \ztp$ in which relations between the elements in $\ztp$ (namely $\mcp^2 = 1$) hold only modulo elements of $\zz_N$. From this perspective, the process of gauging $\ztp$ amounts to ``lifting'' the $\zz_N$ gauge theory up to an $E$ gauge theory. In our case, nontrivial group extensions of $\zz_N$ by $\ztp$ are given by $E = \zz_{2N}$, and so the fact that we obtain a $\zz_{2N}$ gauge theory when the $\omega_\alpha$ are nontrivial is to be expected. On the other hand, if the $\omega_\alpha$ {\it are} trivial, then the $\zz_N$ character of the gauge theory on the mirror axis is unchanged upon gauging $\ztp$, which agrees with the fact that the trivial group extension is given by the choice $E = \zz_N \times \zz_2$. All of these statements also have a natural geometric interpretation in terms of fiber bundles, which we elaborate on in Appendix \ref{sec:bundles}.   

There are several ways of determining whether or not the theory we are working with is anomalous. In what follows, 
we will diagnose possible anomalies by examining the gauge (in)variance of the mirror axis action $S_{\p X}$, with an alternate approach sketched in Appendix \ref{sec:extending_the_fields}. 

A gauge transformation on $S_{\p X}$ means altering the gauge fields $\mcg_\alpha$ by an exact 1-cochain\cite{thorngren2015framed} by sending $\mcg_{\alpha} \mapsto \mcg_{\alpha} + \cob \gamma_{\alpha}$, where $\gamma_\alpha$ is a 0-cochain on $\p X$ taking values in $\zz_2$. We will assume that under this transformation on the $\mcg_\alpha$ the electric and magnetic fields respond by transforming as $a\mapsto a + \lambda_a$, $b\mapsto b+\lambda_b$, where the $\lambda_{\alpha}$ are 1-cochains yet to be determined. Because the symmetry fractionalization pattern of the $a$ and $b$ fields must be the same throughout the entire surface, $a$ and $b$ transform as above on all of $X$, meaning that the action $S_X$ is also affected by the gauge transformation on the $\mcg_\alpha$ fields. 

The total action is generically not gauge-invariant under such a transformation, which signals an anomaly in the theory. We decompose the gauge variance of the total action $S_{tot} = S_X + S_{\mcp(X)} + S_{\p X}$ into three different parts, with the gauge transformation acting on $S_{tot}$ as 
\be S_{tot} \mapsto S_{tot} + \int_{X\cup\mcp(X)}\mco_{top,X}+ \int_{\p X}(\mco_{top,\p X} + \mco_{sym}),\ee 
where the three classes of anomalies are given by  
\be\ba \label{obstructions} & \mco_{top,X} = \frac{1}{N}\left(a\cupp \cob \lambda_b + b\cupp \cob \lambda_a + \lambda_a\cupp \cob \lambda_b\right),\\ 
& \mco_{top,\p X} = \frac{1}{2N}\left((\omega_b \cob \gamma_b -2 \lambda_b) \cupp a + (\omega_a \cob \gamma_a -2\lambda_a)\cupp b\right), \\ 
& \mco_{sym} = \frac{1}{2N}(\omega_b(\mcg_b + \cob \gamma_b)\cupp \lambda_a + \omega_a(\mcg_a + \cob \gamma_a)\cupp \lambda_b \\ & \qquad + 2\lambda_a\cupp \lambda_b).\ea\ee 
In the expression for $\mco_{top,\p X}$, the terms involving the $\lambda_\alpha$ are boundary terms from the gauge variations of $S_X$ and $S_{\mcp(X)}$, which are equal to one another by the requirement of reflection symmetry. 

The ``topological'' anomalies $\mco_{top,X}$ and $\mco_{top,\p X}$ are the most serious of the three. Their nontriviality would imply that the $a$ and $b$ $\zz_N$ gauge fields could not be defined strictly on the (2+1)D surface, and would be instead required to extend into the (3+1)D bulk. Since we rule out bulk topological order by assumption, their contribution $\int_{X+\mcp(X)}\mco_{top,X} + \int_{\p X} \mco_{top,\p X}$ to the gauge variation of $S_{tot}$ must vanish in order for our theory to be well-defined. We see that this happens if we make the choice $\lambda_\alpha = \omega_\alpha \cob \gamma_\alpha/2$, which implies that under the gauge transformation $\mcg_{\alpha} \mapsto \mcg_{\alpha} + \cob \gamma_{\alpha}$ the $\zz_N$ gauge fields transform as
\be \label{eq:abtform} a \mapsto a + \frac{1}{2}\omega_a\cob \gamma_a,\quad b\mapsto b + \frac{1}{2}\omega_b\cob \gamma_b,\ee
which is to be expected and confirms our earlier statement that $\omega_a$ ($\omega_b$) controls the $\ztp$ gauge charge of the quasiparticles in the electric (magnetic) sector. 

The above transformation law \eqref{eq:abtform} for $a$ and $b$ may at first glance look rather trivial, since we should always be free to modify $a$ and $b$ by exact integral 1-cochains without changing the physics of the surface theory. However, $\cob(\omega_{\alpha}\gamma_\alpha/2)$ is only an exact {\it integral} 1-cochain if $\omega_{\alpha}$ is an {\it even} integer. This means that all even choices of $\omega_{\alpha}$ belong to the trivial fractionalization class, while odd choices of $\omega_{\alpha}$ imply nontrivial fractionalization. This can also be checked by explicitly calculating the cohomology group $H^2_\mcp(\zt,\zz_N)$.

This choice for the transformation of the $a$ and $b$ fields eliminates both $\mco_{top,X}$ and $\mco_{top,\p X}$, and so only $\mco_{sym}$ remains, which cannot generically be made to vanish on its own. This represents the gauge anomaly, and is given by the action
\be \label{eq:anomaly_action} S_{anomaly} = \int_{\p X} \mco_{sym} =\frac{\omega_a\omega_b}{4N}\int_{\p X} (\mcg_a \cupp \cob \gamma_b+ \mcg_b\cupp \cob \gamma_a  ),\ee
where we have dropped terms like $d\gamma\cupp \lambda$ and $\lambda\cupp\lambda$ since they are total derivatives. 
Crucially, we see that if either of the electric or magnetic sectors transforms trivially under $\ztp$ then we can set one of the $\omega_\alpha$ to zero, and $S_{anomaly}$ vanishes. That is, if either the electric or magnetic sector transforms trivially under $\ztp$, the theory {\it must} be non-anomalous. This is an illustration of the idea that 
\[ \ba & \textit{anomalies are caused by the interplay between symmetry } \\& \textit{fractionalization in the electric and magnetic sectors. }\ea\] 
Stated differently, having a nontrivial fractionalization class is a necessary, but not sufficient, condition for the occurrence of an anomaly, while having both the electric and magnetic quasiparticles transform projectively under reflection is a sufficient condition. 
This property is closely related to the observation that anomalies which occur in theories with symmetry groups of the form $G_1\times G_2$ arise because of conflicts between the $G_1$ and $G_2$ symmetries\cite{cho2014conflicting,kapustin2014anomalies,kapustin2014anomalousprl}. 

In order for our candidate surface theory to be well-defined, we must cancel the remaining gauge anomaly given by $\mco_{sym}$. It is straightforward to show that the anomalous action $S_{anomaly}$ is the gauge variation of a (2+1)D Chern-Simons-like term, meaning that the anomaly cancellation can be accomplished by the addition of a Chern-Simons action $S_\mcg$ between the electric and magnetic $\ztp$ gauge fields: 
\be \label{eq:mcg_SPT} S_{\mcg} = -\frac{\omega_a\omega_b}{4N}\int_{\Gamma} (\mcg_a \cupp \cob \mcg_b + \mcg_b \cupp \cob\mcg_a),\ee
where $\Gamma$ is some (2+1)D bounding surface with $\p \Gamma = \p X$. This action represents the interaction between the $\mcg_\alpha$ gauge fields, and tells us that anomalies are created when there is a nontrivial topological interaction between the electric and magnetic symmetry fluxes. 

While we expect the anomaly $S_{anomaly}$ to be cancelled by some kind of SPT, we cannot yet be sure whether or not the action $S_\mcg$ represents a genuine SPT phase. 
In order for $S_\mcg$ to correspond to the nontrivial $\zz_2$ SPT, we need to specify the values that the prefactor $\omega_a\omega_b / 4N$ can take. Taking the $\zz_N$ Chern-Simons action \eqref{eq:SX} with $N=2$ initially suggests that the prefactor should be $1/4$. However, the $\mcg_\alpha$ are closed modulo 2, which means that there be no nontrivial quasiparticle excitations (given by the values of the monodromy defects $\cob \mcg_\alpha$ modulo 2) present in the state described by $S_{\mcg}$. This means that since $\cob \mcg_\alpha = 0 \mod 2$, we must multiply the initial factor of $1/4$ by a further factor of $1/2$ if the action is to take on non-integer values\cite{kapustin2014anomalies,thorngren2015higher}. This is analogous to how in the usual $K$-matrix formalism, the difference between $\zz_2$ topological order and the nontrivial $\zz_2$ SPT state is a constant factor of $1/2$ in the $K$-matrix. As a consequence, in order for $S_\mcg$ to represent a nontrivial $\zz_2$ SPT, the prefactor $\omega_a\omega_b/4N$ should actually be equal to $1/8$ (or $0$ if there is no anomaly), which is only true if $\omega_a\omega_b = N/2$. This requirement certainly does not hold for arbitrary symmetry groups, but below we will demonstrate that for $\ztp$ symmetry, these values for $\omega_a\omega_b$ are actually forced upon us. 

A basic physical requirement we can place on the action of reflection symmetry is that it commute with braiding processes. In particular, since braiding processes which exchange two quasiparticles are odd under reflection, we can impose the constraint that $\theta_{\mcp(a),\mcp(b)} = \theta_{a,b}^*$, where $\mcp(\alpha)$ is the image of $\alpha$ under reflection and $\theta_{a,b}$ is the mutual statistics of the quasiparticles associated with the electric and magnetic cochains $a$ and $b$\cite{barkeshli2014symmetry}. Since we restrict ourselves to $\zz_N$ {\it reflection-symmetric} topological order, we can take $\theta_{a,b} = \exp(2\pi ia b /N)$\cite{lin2014generalizations,lake2016signatures}, which means that the action of reflection {\it must} permute the quasiparticles in the system, since it must act as ``charge conjugation'' for either the electric or magnetic sector in order to satisfy $\mcp(a)\mcp(b) = -ab$. 
For concreteness, we adopt the choice that reflection acts as charge conjugation on the magnetic sector, so that $\mcp(a) = a$ and $\mcp(b) = -b$.

These transformation properties can be used to place constraints on the possible reflection quantum numbers of $a$ and $b$ restricted to the mirror axis. Because the vacuum sector must carry a linear representation of $\zz_2^\mcp$, the $\mcp^2$ eigenvalue of an anyon must be the inverse of the $\mcp^2$ eigenvalue of the anyon's dual, as the vacuum must have a $\mcp^2$ eigenvalue of $1$. In terms of the $\omega_\alpha$, this condition is $\omega_\alpha = -\omega_{-\alpha}$. Additionally, acting with $\mcp$ should not change an anyon's reflection quantum number, and so all quasiparticles in a given orbit under $\mcp$ must carry the same reflection quantum number, implying that $\omega_\alpha = \omega_{\mcp(\alpha)}$ for all $\alpha$. Since $\mcp(b) = -b$, we must have $\omega_b=\omega_{\mcp(b)} = \omega_{-b} = -\omega_b$ and so $2\omega_b = 0\ \mod N$, meaning that we are forced to choose $\omega_b\in \{0,N/2\}$ if $N$ is even, and $\omega_b = 0$ if $N$ is odd. We can place no similar restrictions on $\omega_a$, since $a$ is left invariant under reflection. However, we see that in the electric sector, coboundaries are given by $(d\zeta_a)(g,h) = \zeta_{a}(h) - \zeta_a(gh) + \zeta_a(g)$
for $g,h\in \ztp$, and so choosing $\zeta_a(1) = 0,\zeta_a(\mcp) = 1$ tells us that for the electric sector, $2=(d\zeta_a)(\mcp,\mcp) $ is a coboundary in $H^2_\mcp(\zz_2,\zz_N)$. Since coboundaries are gauge degrees of freedom, we are allowed to set $\omega_a \in \{0,1\}$ without loss of generality if $N$ is even, and $\omega_a = 0$ if $N$ is odd. 
 
These choices for $\{ \omega_a,\omega_b\}$ provide us with four fractionalization classes for even $N$, and only the trivial class for odd $N$, which is consistent with the formal calculation of group extensions by ${\rm Ext}(\zz_N^2,\zz_2) = {\rm Ext}(\zz_N,\zz_2)^2 =\zz_{\gcd(2,N)}^2$. The lack of nontrivial fractionalization classes for odd $N$ can also be seen by the fact that cup products in the action $S_{\p X}$ are maps 
\be \wedge : C^1(\p X,\zt) \otimes C^1(\p X,\zz_N) \ra C^2(\p X,\zt\tp \zz_N),\ee 
where $C^k(\p X,G)$ is the set of $k$-cochains on $\p X$ with coefficients in $G$. 
We see that the right-hand side of the above expression must be trivial for $N$ odd, since $\zt\tp\zz_N = \zz_{\gcd(N,2)}$. That is, for odd $N$ there is no way to minimally couple the $a$ and $b$ fields to the $\ztp$ gauge fields, which forces the fractionalization pattern to be trivial. 

Since there are no nontrivial fractionalization classes for odd $N$, we will restrict ourselves to even $N$ in what follows. Additionally, while we focus on the generating fields $a$ and $b$, we should mention that composite quasiparticles which possess both nonzero $\zz_N$ gauge charge and nonzero $\zz_N$ gauge flux may transform under $\mcp^2$ with a possible additional phase factor which depends on the self-statistics of the quasiparticle in question\cite{qi2015detecting,song2016pgspts}. However, this detail will not play an important role in our discussion. 

In passing, we also note that the requirement $\theta_{\mcp(a),\mcp(b)} = \theta_{a,b}^*$ along with  $\theta_{a,b} = \exp(2\pi ia b /N)$ means that when (and only when) $N = 2$, it is also possible for reflection to implement electromagnetic duality by exchanging $a$ and $b$. However, by examining the cocycle relations in the cohomology group $H^2_\mcp(\zz_2,\zz_2)$ we see that $\mfw_a = \mfw_b$, and we also see that coboundaries in $H^2_\mcp(\zz_2,\zz_2)$ are given by 
\be (\cob\zeta_\alpha)(g,h) = \zeta_{\mcp(\alpha)}(h) - \zeta_\alpha(gh) + \zeta_\alpha(g),\ee 
for $\zeta_\alpha : \zt\ra \zz_2$, $\alpha \in \{a,b\}$, and $g,h\in \ztp$. Thus by choosing $\zeta_a(1) = \zeta_b(1) = \zeta_a(\mcp) = 0,\zeta_b(\mcp) = 1$ we have $(\cob \zeta_\alpha)(\mcp,\mcp) = 1$, meaning that $1$ is a coboundary. This allows us to ensure that $\mfw_a=\mfw_b = 0$, implying that the fractionalization pattern must be trivial. Because of this, we will avoid discussing the $\mcp : a\leftrightarrow b$ case in any further detail. 

Since the anomalous action (\ref{eq:anomaly_action}) tells us that the anomaly vanishes whenever $\omega_a$ or $\omega_b$ is trivial, to examine anomalous theories we may set $\omega_a = 1$ and $\omega_b = N/2$ without loss of generality. With this choice, the coefficient in front of the integral in \eqref{eq:mcg_SPT} indeed becomes $1/8$, 
consistent with the interpretation that the mirror axis possesses a $\zz_2$ SPT state
built out of the bosonic $\mcg_\alpha$ gauge fields. Such a $\zz_2$ SPT state is analogous to a Chern-Simons theory with $K$-matrix $K= \sigma^x$\cite{lu2016classification}. 

The final question that remains is the nature of the integration surface $\Gamma$ in (\ref{eq:mcg_SPT}).  One possibility would naively be to simply stack this $\ztp$ SPT state on top of the surface $\zz_N$ topological order, so that $\Gamma = X$. This could be done if we could extend the $\mcg_\alpha$ fields into the interior of the surface $X$ and if the global symmetry acted in an onsite way, but is incompatible with $\ztp$ reflection symmetry, since $\mcp: \cob \mapsto -\cob$ under reflection would imply that $\mcp: S_{\mcg} \mapsto -S_{\mcg}$. We can avoid this issue by regarding $X$ as the surface of a three-dimensional trivial gapped bulk, in which we embed the $\ztp$ SPT state given by $S_{\mcg}$ on the mirror plane by setting $\Gamma = \Sigma$ (see Figure \ref{fig:strings_geometry}). We should stress that we have only gauged reflection symmetry on the mirror axis, not on the entire mirror plane. Since this geometry means that $\ztp$ effectively acts as an onsite symmetry as far as $S_{\mcg}$ is concerned, $S_{\mcg}$ is invariant under reflection (as the action of reflection preserves the $\mcg_\alpha$ fields). 
This choice completes the cancellation of the anomaly, and the final theory is anomaly-free. 

\section{The role of translation symmetry} \label{sec:translation} 
In this section, we briefly comment on how (discrete) translation symmetry enriches our classification of symmetry fractionalization patterns. For simplicity, we will assume that 
the action of translation does not permute the $\zz_N$ gauge fields. 

For an infinite plane geometry, including translation symmetry parallel to the mirror plane (which we will let be implemented by the operator $T_y$) is straightforward. Since no regions in $X$ are mapped to themselves under $T_y$, our dimensional reduction procedure can proceed as in the case with only $\ztp$ symmetry. The full symmetry group of the system is $\ztp \times \zz$, and the Kunneth formula allows us to classify the allowed fractionalization patterns through
\be H^2(\zt\times \zz,\zz_N) = \zz_{\gcd(N,2)}^2.\ee
Note that if reflection symmetry were not present, no symmetry fractionalization could occur (formally, since $H^2(\zz,\zz_N) =0$). Thus, the presence of translation symmetry enriches the full classification, even though translation symmetry cannot contribute any nontrivial fractionalization classes on its own. 

The situation is changed if $X$ has periodic boundary conditions in the $y$ direction.
For example, if $X$ is a cylinder such that the mirror axis is an $S^1$ of length $L$ (in units of the lattice constant), the full symmetry group of the system is changed to $\ztp \times \zz_L$. The dimensional reduction procedure proceeds normally, but the number of fractionalization patterns is changed. In this case, the Kunneth formula gives 
\be H^2(\zt\times\zz_L,\zz_N) = \zz_{\gcd(N,2)} \times \zz_{\gcd(L,2)} \times \zz_{\gcd(N,L,2)},\ee
where we recognize the three factors as the fractionalization of reflection, translation, and the interplay between the two (which is the fractionalization of the relation $(T_y\mcp)^L = 1$ if $L$ is even). 

Translation normal to the mirror plane (generated by the operator $T_x$) can be dealt with by following the procedure of Ref\cite{song2016pgspts}.  The symmetry group relation between reflection and translation is $\mcp T_x \mcp = T^{-1}_x$ or equivalently $(\mcp')^2=1$, where $\mcp' = \mcp T_x$ is an operator which reflects about a plane located at $x=1/2$. The enumeration of the fractionalization classes proceeds for the $\mcp'$ symmetry in the same way as it did for the $\mcp$ symmetry, and so the classification of fractionalization patterns is enlarged to $\zt^4$, with one factor of $\zt^2$ coming from the four fractionalization classes associated with $\mcp$ and the other coming from the four classes associated with $\mcp'$. Applying the field theory construction of the previous section shows that anomalies occur when both $a$ and $b$ fields transform projectively under the action of either $\mcp$ or $\mcp'$. For anomalous $\mcp$ ($\mcp'$) fractionalization patterns that neccesitate the presence of a (3+1)D bulk geometry, a stack of $\zt$ SPT phases captured by the action \eqref{eq:anomaly_action} will be forced to live on each $\mcp$ ($\mcp'$) mirror plane, as argued in Ref\cite{song2016pgspts}.

\section{Discussion} \label{sec:disc}

To summarize, we have shown that for (2+1)D $\zz_N$ topological phases enriched with reflection symmetry, a 
dimensional reduction approach can be applied to obtain a concrete classification of all anomalous and non-anomalous symmetry fractionalization patterns. The anomalies that do appear can be cancelled with a (2+1)D (rather than (3+1)D) $\zz_2$ SPT state built out of a pair of $\zz_2$ gauge fields, as long as the SPT comes embedded in an otherwise trivial ambient (3+1)D bulk. Thus, although the presence of a three-dimensional bulk is still crucial for making the full theory well-defined, its existence plays a rather passive role in the anomaly cancellation. 

Furthermore, we showed that the anomalous fractionalization patterns are those where both the electric and magnetic quasiparticles transform projectively under reflection, with anomalies occurring when the electric and magnetic $\ztp$ gauge fields introduced during the anomaly detection procedure interact with each other in a nontrivial way. This means that if either the electric or magnetic sector transforms under a linear representation of $\ztp$ the theory is {\it guaranteed} to be non-anomalous, regardless of what happens in the other sector. For the $\zz_N$ topological order we have focused on, we found that for odd $N$ there are no nontrivial fractionalization classes, while for even $N$ there are always four, corresponding to the trivial class, the class in which the electric sector transforms projectively, the class in which the magnetic sector transforms projectively, and the class in which both sectors transform projectively. Our results imply that the last class is always anomalous, and that the first three never are. Several recent works have studied the example of $N=2$ from different perspectives, and their results agree with ours in this case\cite{qi2015anomalous,song2016pgspts}. This idea also agrees with results concerning anomalous fractionalization patterns in similar theories enriched with $\ztt$ time reversal symmetry \cite{peng2015vortex}. 

While we have only focused on reflection and translation symmetry in (2+1)D $\zz_N$ topological phases, the extension to other space group symmetries, different types of abelian topological orders, and systems with different spacetime dimensions is straightforward. Incorporating larger symmetry groups of the form $G_{tot} = G_{space} \times G_{spin}$ for a Lie group $G_{spin}$ can also be done fairly straightforwardly. 
Besides anomalies resulting from anomalous $G_{space}$ and $G_{spin}$ fractionalization patterns, there will generically be additional anomalies occurring from the interaction between the $G_{space}$ and $G_{spin}$ gauge fields. For example, performing our analysis for $G_{space} = \ztp$ and $G_{spin} = U(1)$ or $SO(3)$ shows that if a quasiparticle carries $\mcp^2 = -1$, an anomaly will occur if the particle's electromagnetic dual carries half-odd-integer spin or half-odd-integer charge, which has been noticed previously for the example of $\zz_2$ topological order \cite{qi2015anomalous,hermele2015bosonic}.

The extension to more general types of abelian topological order is also straightforward, and simply involves performing the analysis described above for theories with different reflection-symmetric Chern-Simons terms in the action for $S_X$. This is simplified by the fact that the requirement of having a reflection-symmetric action is a rather strong one, since the basic Chern-Simons term $a \wedge \cob a$ is odd under reflection. Furthermore, even reflection-symmetric actions are often forced to have trivial fractionalization patterns. For example, consider the Lagrangian $\mcl_X = (a\wedge da - b\wedge db)/(2N)$, which is symmetric under reflection provided that reflection interchanges the $a$ and $b$ fields. However, as mentioned in the previous section, $\mcp : a \leftrightarrow b$ implies that $H^2_\mcp(\zt,\zz_N) = 0$, meaning that there are no nontrivial fractionalization classes in such a theory. Additionally, following our procedure for actions with terms like $a\wedge b \wedge c$ suggests that each of the fields $a,b,c$ must carry a trivial reflection quantum number, as the cancellation of $\mco_{top,X}$ seems to require $\lambda_a=\lambda_b=\lambda_c=0 \mod N$. It would be interesting to explore these examples further and to determine whether or not non-abelian topological phases can be treated in a similar way. 



\acknowledgements
I am very grateful to Michael Hermele, Ryan Thorngren, and the participants of the Boulder School for Condensed Matter and Materials Physics for many helpful discussions. I am also indebted to Judith Hoeller for useful comments on the manuscript. This work was partially supported by the NSF grant PHY-1560023.

\appendix

\section{Review of cochains and cup products} \label{sec:algtop}

Our discussion in the main text and below draws heavily on the notion of cochains and cohomology groups, and so in this section we provide a brief review of the essential definitions\cite{bott2013differential,brown2012cohomology}. We define a {\it simplicial $k$-cochain $A$ on $X$ with coefficients in $M$} to be a function $A$ from the $k$-simplicies of a triangulated manifold $X$ to the ring $M$, which we typically take to be $M = \zz$ or $M =\zz_N$ (recall that a 0-simplex is a vertex, a 1-simplex is a link, a 2-simplex is a face, and so on). We write $A \in C^k(X,M)$ to denote such a cochain, which is the discrete analogue of a differential $k$-form. Cochains come with a natural map $\cob$ called the coboundary operator, which increases their degree by 1 ($i.e.$ $\cob : C^k(X,M) \ra C^{k+1}(X,M)$), and is defined by summing up the values that the cochain takes around closed loops in the triangulation of $X$. As the notation suggests, the coboundary operator $\cob$ is the discrete analogue of the exterior derivative. 
Explicitly, for $A\in C^k(X,M)$, we have 
\be (\cob A)(p_0,\dots,p_{k+1}) = \sum_{i = 0}^{k+1}(-1)^{i}A(p_0,\dots,\hat p_i,\dots,p_{k+1}),\ee
where the $p_i$ are points in $X$ and $\hat p_i$ indicates that the entry $p_i$ is omitted. The notation $A(p_0,\dots,p_k)$ is taken to mean the function $A$ evaluated on the $k$-cell in $X$ formed by the points $p_0,\dots,p_k$. 
As in exterior calculus, $k$-cochains $A = \cob \alpha$ for $\alpha \in C^{k-1}(X,M)$ which are coboundaries of $(k-1)$ cochains are called exact, and typically represent gauge degrees of freedom. Cochains with zero coboundary ({\it i.e.} those satisfying $\cob A = 0$) are called cocycles. The group of all $k$-cocycles modulo the exact ones is called the $k$-th simplicial cohomology group of $X$ with coefficient ring $M$, and is written as $H^k(X,M)$. 

In many applications, like when determining the $\mfw_\alpha$ cocycles used in the main text, $X = G$ will actually be a finite group, rather than a manifold. In this case, a cochain $A \in C^k(G,M)$ is best viewed as a function $A : G^k \ra M$. When the group $G$ is equipped with a nontrivial action $\rho$ on $M$, the cohomology group is ``twisted'' by this action. The coboundary operator $\cob$ is also twisted, and for $A\in C^k(G,M)$, $\cob A$ takes on the explicit form
\begin{widetext}
\be (\cob A)(g_1,\dots,g_{k+1})  = \rho_{g_1}[A(g_2,\dots,g_{k+1})] + \sum_{i=1}^{k}(-1)^i A(g_1,\dots,g_{i-1},g_{i}g_{i+1},g_{i+2},\dots,g_{k+1})  + (-1)^{k+1}A(g_1,\dots,g_k),\ee
\end{widetext}
where we have used additive notation for $M$ and multiplicative notation for $G$. For example, for $A\in C^1(G,M)$ we have 
\be (\cob A)(g_1,g_2) = \rho_{g_1}[A(g_2)] - A(g_1g_2) + A(g_1)\ee 
and for $A\in C^2(G,M)$ we have 
\be \ba (\cob A)(g_1,g_2,g_3) = \rho_{g_1}&[A(g_2,g_3)] - A(g_1g_2,g_3) \\ &+ A(g_1,g_2g_3) - A(g_1,g_2).\ea \ee 
As in the simplicial case, we can define a cohomology group $H^k(G,M)$ consisting of all closed $k$-cochains modulo the exact ones. The notions of group cohomology and simplicial cohomology are connected through a space $BG$ known as the classifying space of $G$, which has $\pi_1(BG) = G, \pi_{i>1}(BG)=0$, and is such that the simplicial cohomology $H^k(BG,M)$ is the same as the group cohomology $H^k(G,M)$. Simple examples relevant for us are $B\zz = S^1$ and $B\zt = {\mathbb{RP}}^\infty$. 

\begin{figure}
\includegraphics[scale=0.8]{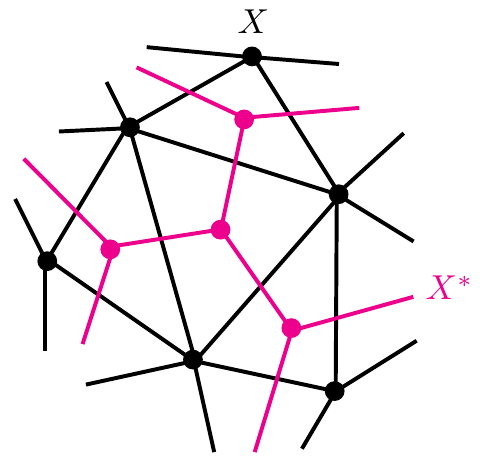}
\caption{\label{fig:poincare_duality} An illustration of Poincare duality for a region of a 2-dimensional triangulated manifold $X$, shown in black. The poincare dual of $X$ is shown in pink.}
\end{figure}

When studying symmetry fractionalization in topological field theories, simplicial cochains on a triangulated manifold $X$ are often viewed algebraically as 
functions on the cells of $X$. We feel that the dual formulation we take in this paper
can also be very useful: instead of thinking of cochains as functions on $X$, we view them as submanifolds of $X$, adopting a geometrically-focused
perspective. Our ability to do this is based on the notion of Poincare duality, which for us means that any 
$k$-cochain $A$ is dual to an $(n-k)$-dimensional submanifold of $X$, where $n = \dim X$. On triangulated manifolds, this
is illustrated by the fact that the Poincare dual of a cochain $A$ is simply given by the form the cochain takes on the dual graph of $X$. An example of this is shown in Figure~\ref{fig:poincare_duality}, where the black graph represents a region of $X$ and the pink graph represents a region of $X^*$, the poincare dual of $X$. In the figure $\dim X = 2$, and so the dual of a 1-cochain is a 1-manifold, while the dual of a 0-cochain is a 2-manifold, and vice versa. From the figure, it is clear that a 1-cochain defined on $X$ (like $a$ from the main text) always intersects a 1-cochain defined on $X^*$ (like $b$ from the main text) transversely. 

With this in mind, we see that in the surface theory \eqref{eq:SX} for $X$, the 1-cochain $a$ is dual to a 2-dimensional submanifold $a^* \subset X$ which represents the worldsheets swept out by the electric $a$ strings, and likewise for $b$. On the other hand, the 2-cochain $\cob a$ is dual to a 1-dimensional submanifold $(\cob a)^* \subset X$, which represents the worldlines traced out by the spacetime trajectories of the electric quasiparticles, and similarly for $(\cob b)^*$. This is because the coboundary operator becomes the boundary operator $\p$ under Poincare duality, and so $(\cob a)^* = \p a^*$ corresponds to the ends of electric strings, which are identified with electric quasiparticles. In the topological field theories we are interested in, electric and magnetic quasiparticles interact with each other by passing through each other's strings. Therefore, we expect that a field theory description of such topological order should capture the intersections between $a^*$ and $(\cob b)^*$ and those between $b^*$ and $(\cob a)^*$. 

In order to more precisely incorporate interactions into our field theory construction, we would like to be able to ``multiply'' two cochains together. This is done by using a device known as the {\it cup product}, which takes a $k$ cochain $A$ and an $l$ cochain $B$ to a $k+l$ cochain. We will abuse notation slightly and write $A\cupp B$ for the cup product of $A$ and $B$ (the usual notation is $\cup$). Algebraically, the cup product has a very simple definition:
\be (A \cupp B)(p_0,\dots,p_{k+l}) = A(p_0,\dots,p_k)B(p_k,\dots,p_{k+l}),\ee
where as before, the $p_i$ are points in the spacetime manifold $X$.   
Like the regular wedge product, a key algebraic feature of the cup product is its supercommutativity, namely that for any $k$-cochain $A$ and any $l$-cochain $B$, we have 
\be \label{eq:cup_ids} \ba &A \cupp B = (-1)^{kl}B\cupp A,\\  &\cob(A\cupp B) = \cob A\cupp B+ (-1)^kA\cupp\cob B.\ea\ee

Instead of thinking about the cup product algebraically, we prefer to invoke Poincare duality to establish a geometric, rather than algebraic, interpretation of $A\cupp B$. Under Poincare duality, 
the multiplication of functions becomes the intersection of submanifolds, and so $A \cupp B$ can be interpreted as the intersection between the two submanifolds represented by $A$ and $B$. That is, we have
\be \label{eq:cupcap} A \cupp B = A^* \cap B^*.\ee
The intersection $\cap$ is oriented, so that the cup product keeps track of the relative orientation with which two manifolds intersect. 
The intersection is also defined modulo homotopy equivalence, so that the RHS of \eqref{eq:cupcap} denotes the homology class of $A^* \cap B^*$. 

\begin{figure}
\includegraphics[scale=.8]{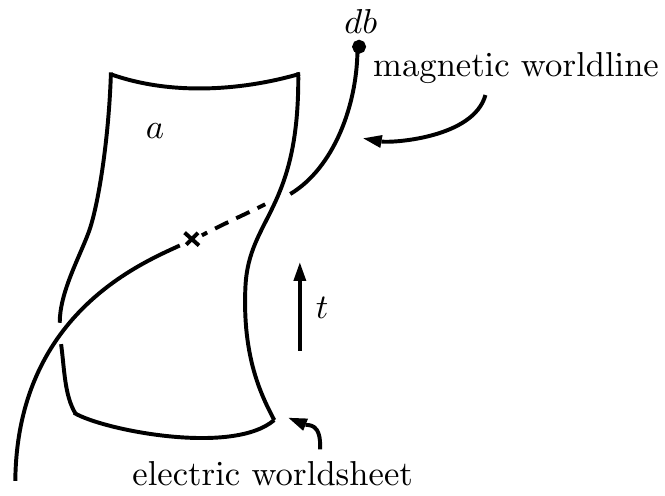}\caption{\label{fig:cupfig}The geometric meaning of the term $a\cupp \cob b$. Since $a$ ($\cob b$) is a 1-cochain (2-cochain), it is represented by a codimension 1 (2) submanifold in (2+1)D spacetime. The intersection of these submanifolds (marked by the x) is a codimension 3 submanifold, and is represented by the cup product $a\cupp\cob b$. }
\end{figure}

We are now ready to better motivate our action $S_X$ for the surface $\zz_N$ topological order, which has the familiar Chern-Simons form
\be \label{eq:SXapp} S_{X} = \frac{1}{2N}\int_X(a\cupp \cob b + b\cupp \cob a).\ee
In the action, the term $a\cupp \cob b$ corresponds to the intersection mod homotopy of $a$-string worldsheets and the boundaries of $b$-string worldsheets, which correspond to magnetic quasiparticles. The action simply tells us that when magnetic quasiparticles pass through electric worldsheets, they pick up a nontrivial braiding factor. Integrating $a\cupp \cob b$ by parts gives $\cob a \cupp b$ (on a closed manifold), and so we can equivalently interpret the braiding process represented by $a\cupp \cob b$ as {\it electric} quasiparticles passing through {\it magnetic} worldsheets, which corresponds to the second term in \eqref{eq:SXapp}. Alternatively, we can think of $b$ as a Lagrange multiplier field ensuring the flatness of $a$, and vice versa.
Figure \ref{fig:cupfig} provides a pictorial illustration of the $a\cupp \cob b$ term. 

The notion of cup products and coboundary operators is very similar to the concepts of their more familiar differential geometry twins, with the cup product serving as the discrete analogue of the 
wedge product and the coboundary operator serving as the discrete analogue of 
the exterior derivative. 
The reader may then wonder why we bother using cup products instead of differential forms, if they are so closely related to one another. The main advantage of working with cup products and cochains rather than wedge products and differential forms is that cup products allow us to more naturally incorporate the algebraic information of the symmetry fractionalization class (like the factor sets $\mathfrak{w}_\alpha$) into the theory at a field-theoretic level. Additionally, working with cochains allows our field theory to be naturally defined on discrete lattices with integer-valued cochains, which we feel is more natural when studying topological orders constructed from finite groups. 


\section{A geometric interpretation of symmetry fractionalization}\label{sec:bundles}

In this section, we briefly outline a geometric interpretation of symmetry fractionalization, which is helpful for understanding the physical meanings of the algebraic objects like the factor sets $\mfw_\alpha$ used in the main text. In this section, we will work in a more general setting in which an abelian topological order derived from a finite group $N$ is enriched by a symmetry group $G$, which need not be finite.

As mentioned in the main text, the possible fractionalized quantum numbers for the electric and magnetic sectors of the theory are parametrized by a choice of cohomology class $\mfw_\alpha \in H^2_\rho(G,N)$, where the cohomology group is twisted by the action $\rho$ of $G$ on $N$ and $\alpha \in \{a,b\}$ distinguishes between the electric and magnetic sectors. The set of possible choices for $\mfw_\alpha$ contains the same information as the set of (not necessarily central) group extensions of $G$ by $N$, which in turn correspond to the different ways of constructing exact sequences 
\be \label{sequence}1 \ra N \ra E \ra G \ra 1.\ee
Trivial fractionalization classes $\mfw_\alpha = 0$ correspond to split sequences in which $E$ is given by a semi-direct product $E = N \rtimes G$, while nontrivial fractionalization classes correspond to scenarios in which $E$ is not a product group. For example, if we take $N=\zz_N$ and $G=\zz_2$ with the trivial action of $G$ on $N$, then the trivial fractionalization class corresponds to the choice $E = \zz_N\times \zz_2$, with the nontrivial choice corresponding to $E = \zz_{2N}$. 

\begin{figure}
\includegraphics[scale=.8]{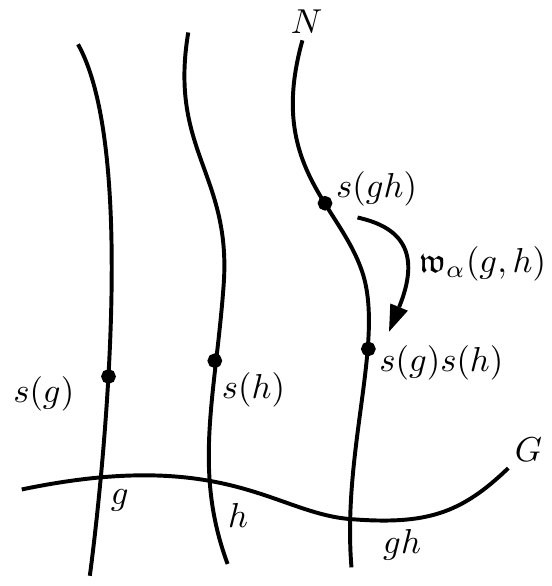} 
\caption{ \label{fig:proj_bundle}What fractionalization looks like from a geometric perspective. Here the fiber $N$ describes the topological phase, and the base space $G$ describes the symmetry group. $s$ is a section (aka projective representation) which lifts $G$ up into the fibers. The fractionalization factor set $\mfw_\alpha$ measures the failure of $s$ to be linear.  }
\end{figure}

While group extensions are usually thought of as algebraic objects, they have a very natural geometric interpretation in terms of fiber bundles. In this interpretation, $E$ is a fiber bundle built out of the base space $G$ and fibers given by the group $N$.\footnote{Readers uncomfortable working with fibers and base spaces built from discrete groups may replace all finite groups by their classifying spaces and the exact sequence \eqref{sequence} by the associated fibration $BN \ra BE \ra BG$.}
In the fiber bundle picture, representations of the symmetry group become maps from the base space up into $E$. 
Let us denote the representations of the symmetry group $G$ by $s: G \ra E$, which is the same thing as a section of the bundle $E$. That is, the representation $s : G \ra E$ ``lifts'' up the base space $G$ into the fibers. If $s$ is a linear representation, it must take values in the image of $N$ under the map $N \hookrightarrow E$. However, allowing $s$ to be a projective representation means that $s$ may actually take on arbitrary values in the group $E$.  $s$ is a projective representation when it fails to be a homomorphism, which is the same as saying that if fails to do the ``lifting'' of $G$ in to $E$ in a linear way. If this is the case, $s(g)s(h)$ is only equal to $s(gh)$ up to a phase factor, which we define as $\mfw_\alpha(g,h)$, with $\alpha \in \{a,b\}$ specifying whether we are working in the electric or magnetic sector. The phases $\mfw_\alpha(g,h)$ are needed to ``untwist'' movement around in the bundle, and are precisely the algebraic objects used in the main text to designate the symmetry fractionalization pattern. These concepts are illustrated in Figure \ref{fig:proj_bundle}.

We can now obtain a clearer understanding of the mirror axis action \eqref{eq:spx}, which can be easily generalized to the more general case we are focusing on in this section. $\mfw_\alpha$ measures the amount by which a given position in a fiber changes as the base point of the fiber is moved around a loop in the base space, and forming the cochain $\mcg_\alpha^*\mfw_\alpha$ (which is equal to $\omega_\alpha \mcg_\alpha/2$ in the context of the main text) by pulling back $\mfw_\alpha$ by $\mcg_\alpha : X \ra G$ translates the curvature in the bundle $E$ with base space $G$ into the curvature of the pullback bundle $\mcg_\alpha^*E$ with base space $X$. If there is an anomaly, the curvature in the fibers caused by $\mfw_\alpha$ cannot be ``untwisted'', which creates an obstruction to forming a complete $E$ gauge theory. The anomaly is canceled by coupling the $N$ gauge fields to the $\mcg_\alpha$ fields by writing an action with terms like $\alpha\cupp\mcg_\alpha^*\mfw_\alpha $, which ``equivariantizes'' movement in the fibers and promotes the $N$ gauge theory to an $E$ gauge theory, gauging $G$ in the process. Geometrically, this procedure can be visualized by thinking of the initial $N$ gauge theory as restricted to live within a single fiber, marked by the base point of the identity $1\in G$. Gauging $G$ allows us to hop from fiber to fiber in the full bundle, and the result is a full $E$ gauge theory. Our results from the main text tell us that for $G = \ztp$ and $N = \zz_N$, the theory is anomalous whenever both the electric and magnetic bundles are twisted in a nontrivial way. 

Finally, we point out that the twisting in the fiber bundle caused by $\mfw_\alpha$ manifests itself algebraically as a ``perturbation'' to the group multiplication law in $E$. If we write elements in $E$ as pairs $(a,g)$ where $a\in N$ and $g\in G$ and let $\rho$ denote the action of $G$ on $N$, then we can capture the full structure of the bundle by working with the following group multiplication law in $E$:
\be (a,g) (b,h) = (a + \rho_g( b) + \mfw_\alpha(g,h),gh).\ee
This is essentially a perturbed version of a regular semi-direct product action, with the factor set $\mfw_\alpha$ acting as the perturbation to the semi-direct product structure. The fiber bundle interpretation shows how this algebraic perturbation manifests itself as a twist in a geometric way.

\section{An alternate calculation of the anomaly} \label{sec:extending_the_fields}

In the main text, we computed the anomaly by examining the gauge variation of the mirror axis action $S_{\p X}$. In this section, we present an equivalent but more geometrically-minded calculation of the anomaly. Instead of testing the gauge invariance of the action, we test its ability to be defined in strictly (1+1)D by asking whether or not it contains any hidden information about (2+1)D physics. 

The basic idea \cite{dijkgraaf1990topological,kapustin2014anomalies} is to compute the curvature of the mirror-axis Lagrangian, $\cob \mcl_{\p X}$, to see if the theory on $\p X$ is twisted in some fundamental way that only makes sense in the presence of higher-dimensional fields. In general, if $\mcl$ is a $k$-cochain on a $k$-dimensional manifold $M$, then we must have $\cob \mcl = 0$. To see this, note that by Poincare duality we may associate $\mcl$ with a zero-dimensional submanifold of $M$, and because Poincare duality maps $\cob \mapsto \partial$ and zero-dimensional submanifolds always have zero boundary, we must have $\p (\mcl)^* = 0$ and hence $\cob \mcl = 0$. 
Applying this to the problem at hand, we see that if $\cob \mcl_{\p X}$ is nontrivial, then the action only makes sense when the fields are extended to a (2+1)D manifold $\Gamma$ with $\p \Gamma = \p X$, and the action $S_{\p X}$ is anomalous. 

As in the main text, we take 
\be \mcl_{\p X} = \frac{1}{2N}( \omega_b \mcg_b \cupp a +\omega_a\mcg_a \cupp b)\ee 
to represent the mirror axis Lagrangian after we have attempted to gauge $\ztp$ on the mirror axis. 
We write the coboundary of the Lagrangian as $\cob \mcl_{\p X} = \mco_{top} + \mco_{sym}$, where the anomaly is split into a topological part $\mco_{top}$ which explicitly involves the gauge fields $a$ and $b$, and a symmetry-related part $\mco_{sym}$. These two classes of anomalies are given by 
\be \ba \mco_{top} & = \frac{1}{2N}(a\cupp \omega_b \cob\mcg_b  + b\cupp\omega_a\cob\mcg_a ),\\
\mco_{sym} & = -\frac{1}{2N}(\cob a \cupp \omega_b \mcg_b + \cob b \cupp \omega_a\mcg_a).\ea\ee

As before, the ``topological'' part of the anomaly $\mco_{top}$ is the more severe of the two, and its nontriviality would imply a fundamentally ill-defined theory. This requirement would naively imply that we must set $\omega_a = \omega_b = 0$, which would force our theory to have trivial fractionalization. However, we note that we can identify $\mco_{top}$ with the surface topological order Lagnrangian $\mcl_X + \mcl_{\mcp(X)}$ provided that we place the following constraints on the curvatures of the $\zz_N$ gauge fields:
\be \label{eq:deladelb} \cob a = \frac{1}{2}\omega_a \cob\mcg_a,\quad \cob b = \frac{1}{2}\omega_b\cob\mcg_b.\ee

This requirement on the curvatures of the $\zz_N$ gauge fields is actually a very natural one. We know that $\cob a$ ($\cob b$) measures the integral of electric (magnetic) strings about closed loops in the spacetime manifold, and so in a ground state with no quasiparticle excitations present, the flatness constraints $\cob a = \cob b = 0$ will be modified only by the presence of symmetry fluxes, which are monodromy defects that introduce curvature into the $\zz_N$ gauge fields. The worldines of the symmetry fluxes are given explicitly by the cochain $\mcg_\alpha^*\mfw_\alpha$, the pullback of $\mfw_\alpha$ by the gauge field $\mcg_\alpha : X \ra \zz_2$ \cite{kapustin2014anomalies}. Since the curvatures $da$ and $db$ will depend only on locations of the symmetry flux worldlines, we expect that $\cob a = \mcg^*_a\mfw_a$ and $\cob b = \mcg_b^*\mfw_b$. 
At the same time, we can set $\mfw_\alpha(g,h) = 0$ if either of $g$ or $h$ is the identity, and so the only nontrivial value of $\mfw_\alpha$ is $\mfw_\alpha(\mcp,\mcp) = \omega_\alpha$. Thus, the symmetry flux worldlines will be described by a 2-cochain which takes the value $\omega_\alpha$ in locations where the gauge field $\mcg_\alpha$ takes on nontrivial values, and which vanishes on locations where $\mcg_\alpha$ vanishes. This is equivalent to making the identification $\mcg^*_\alpha\mfw_\alpha = \omega_\alpha \cob \mcg_\alpha/2$, with the factor of $1/2$ coming from the fact that $\mcg_\alpha$ is closed modulo $2$. Using $\cob \alpha = \mcg^*_\alpha \mfw_\alpha$, we recover the constraints \eqref{eq:deladelb}. 


With $\mco_{top}$ cancelled by introducing nontrivial curvatures for the $a$ and $b$ fields, all that remains is to cancel the gauge anomaly $\mco_{sym}$. 
Plugging the relations \eqref{eq:deladelb} into the expression for $\mco_{sym}$, we obtain the following action for the anomaly:
\be \label{eq:sanomalyapp} S_{anomaly} = -\frac{\omega_a\omega_b}{4N}\int_\Gamma(\mcg_a\cupp\cob\mcg_b + \mcg_b\cupp\cob\mcg_a).\ee
This tells us that in order for the theory to be well defined, we must have a $\zz_2$ SPT state represented by \eqref{eq:sanomalyapp} present on some bounding manifold $\Gamma$ with $\p \Gamma = \p X$. As explained in the main text, reflection symmetry forces us to set $\Gamma = \Sigma$ to be the mirror plane, which completes the cancellation of the anomaly.

\section{Motivating dimensional reduction with the folding trick}\label{sec:folding}

\begin{figure}
\includegraphics[scale=.8]{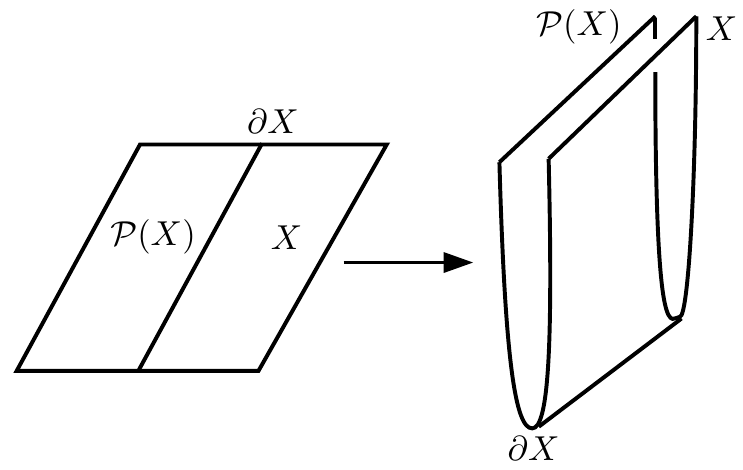} 
\caption{ \label{fig:foldingtrick} An illustration of the folding trick. We fold the surface along the mirror axis $\p X$, so that the system becomes a bilayer system with edge $\p X$ on which reflection acts as a layer-exchange symmetry.  }
\end{figure}

In this section we provide another (very schematic) way to motivate our dimensional reduction approach, inspired by the folding trick \cite{kitaev2012models,lan2015gapped}. Focusing only on the surface theory, we note that we can fold up the surface about the mirror axis while preserving reflection symmetry, until we obtain a geometry in which reflection effectively acts as an ``onsite'' layer-exchange symmetry on a doubled sheet carrying two mirror-symmetric copies of the surface action \eqref{eq:SX}, whose boundary is the mirror axis $\p X$ (see Figure \ref{fig:foldingtrick}). Let us denote the doubled sheet obtained after folding by $X\boxtimes \mcp(X)$. Using the antisymmetry properties of the cup product, one can integrate by parts to show that 
\be \ba S_{X\boxtimes \mcp(X)} & = \frac{1}{N} \int_{X\boxtimes \mcp(X)} (a\cupp \cob b + b\cupp \cob a) \\ & = \frac{2}{N}\int_{X\boxtimes \mcp(X)} b\cupp \cob a - \frac{1}{N}\int_{\p X} a\cupp b.\ea\ee
On the other hand, we can also integrate by parts to get 
\be \ba S_{X\boxtimes \mcp(X)} = \frac{2}{N}\int_{X\boxtimes \mcp(X)} a\cupp \cob b + \frac{1}{N}\int_{\p X} a\cupp b .\ea \ee
Combining these two equations tells us that 
\be \label{eq:foldaction} \frac{1}{N}\int_{\p X} a \cupp b = \frac{1}{N} \int_{X\boxtimes \mcp(X)}( - a\cupp \cob b+b\cupp \cob a).\ee

As explained in the main text, reflection symmetry forces either $\mfw_a$ or $\mfw_b$ to take values only in $\{0,N/2\}$. For concreteness, we choose that this constraint be imposed on $\mfw_b$. We can then imagine introducing symmetry fluxes for reflection on the mirror axis by way of the gauge fields $\mcg_a$ and $\mcg_b$ defined in the main text. As explained in Appendix \ref{sec:extending_the_fields}, symmetry fluxes are monodromy defects for the $\zz_N$ gauge fields, and are responsible for modified flatness constraints $\cob a = \mcg^*_a\mfw_a, \cob b = \mcg^*_b\mfw_b$ on the $a$ and $b$ gauge fields. In particular, this means that we must have $2\cob b = 0\mod N$, since $\mfw_b$ must always take values in integer multiples of $N/2$. This means that the right-hand side of \eqref{eq:foldaction} is actually {\it equal} to $S_{X\boxtimes \mcp(X)}$, since only values of $S_{X\boxtimes \mcp(X)}$ mod $1$ are physical and the coefficient in front of the $a\cupp \cob b$ term after using using the restriction on $\mfw_b$ becomes $-1/2 = 1/2$ mod $1$. This means that the surface theory can be dimensionally reduced to a (1+1)D theory defined only on the mirror axis $\p X$. In this way, we see that the constraints imposed by reflection symmetry allow us to write the surface theory {\it entirely} in terms of an action defined on the mirror axis, and so to study fractionalization and anomalies in these theories we expect to be able to focus solely on the behavior of the simpler (1+1)D physics of the mirror axis, which motivates the dimensional reduction approach considered in the main text.

\bibliography{reflection_SETs_paper}

\end{document}